\documentclass[longbibliography, twocolumn, amsmath, amssymb, aps, superscriptaddress]{revtex4-2}
\usepackage{xr}
\usepackage{amsmath}
\usepackage[utf8]{inputenc}
\usepackage{graphicx}   
\usepackage{dcolumn}  
\usepackage{bm}          
\usepackage{xcolor}    
\usepackage{verbatim}
\usepackage{natbib}

\usepackage{hyperref}
\hypersetup{
    colorlinks=true,
    linkcolor=blue,
    filecolor=black,      
    urlcolor=blue,
    citecolor=blue,
}
\begin{document}

\title{Spontaneous Motion of Liquid Droplets on Soft Gradient Surfaces}
\author{Weiwei Zhao}
\thanks{These authors contributed equally}
\affiliation{Department of Physics, The Hong Kong University of Science and Technology, Hong Kong, China}
\author{Wenjie Qian}
\thanks{These authors contributed equally}
\affiliation{Department of Physics, The Hong Kong University of Science and Technology, Hong Kong, China}
\author{Chang Xu}
\affiliation{Department of Physics, The Hong Kong University of Science and Technology, Hong Kong, China}
\author{Qin Xu}
\email[]{qinxu@ust.hk}
\affiliation{Department of Physics, The Hong Kong University of Science and Technology, Hong Kong, China}

\begin{abstract}
We report an experimental investigation of the spontaneous motion of liquid droplets on soft gels with a crosslinking gradient. By systematically adjusting the spatial difference in crosslinking density, we observed that millimeter-sized liquid droplets moved along the  elastic modulus gradient and even climbed inclined slopes against gravity. Unlike the wetting dynamics of micro-droplets, which are governed by elastocapillary effects, we demonstrated that the observed spontaneous movements of millimeter-sized  droplets were driven by the surface energy difference resulting from the variations in crosslinking density. Using {\em in-situ} confocal microscopy imaging, we analyzed the viscoelastic dissipation induced by the moving wetting ridges near dynamic contact lines. Our findings provide a novel strategy for controlling droplet dynamics on soft and dissipative interfaces, based on the relationship between crosslinking density and surface energy of soft gels.   

\end{abstract}
\date{\today}
\maketitle

\section{Introduction}

The dynamic wetting of droplets on soft interfaces plays a crucial role in various biological processes and engineering applications, including cellular mechanosensing~\cite{Trichet2012}, tissue growth on soft substrates~\cite{Discher2005},  microfluidic device design~\cite{Cubaud2008}, and micro-scale drug delivery~\cite{Yuan2010}. Moreover, moving liquid droplets have been used as probes to characterize the tribology and interfacial rheology of soft interfaces~\cite{Lhermerout2016, Gao2018, Khattak2022}. Therefore, precise control of droplet motion on soft surfaces is highly desirable from both scientific and engineering perspectives. 

However, the manipulation of droplet dynamics on compliant surfaces remains less explored than that on rigid substrates~\cite{Wong2011, Chaudhury2011, Lafuma2003, Courbin2007, Xu2012}. Conventional methods, such as chemical deposition and surface roughness patterning~\cite{Coux2020, Dadhichi2014}, faces significant challenges in modifying the surface energy of soft surfaces due to the presence of diffusive solvents and solid capillarity. Although theoretical models have been used to predict droplet motion on soft gradient surfaces~\cite{Bardall2020, bueno2018wettability,theodorakis2017stiffness, Leong2020}, direct experimental validation of these predictions remains limited.  Research has shown that micro-droplets can spontaneously move along thickness gradients on soft gels due to elastocapillary effects~\cite{style2013patterning}. However, this spontaneous movement occurs only when the droplet sizes are comparable to the elastocapillary length of the substrates, typically of the order of $10$~$\mu$m. 

In this work, we experimentally observe of the spontaneous motion of millimeter-sized droplets on soft gel surfaces with sharp crosslinking gradients. Through measurements of dynamic contact angles and confocal imaging of moving contact ridges, we demonstrate that the difference in crosslinking density leads to variations in the surface energy of soft gels. The resulting imbalance in contact forces can cause  large liquid droplets to move spontaneously, thereby counterbalancing the viscoelastic dissipation from substrates and even overcoming gravity along inclined surfaces.

\section{Experimental System}
\subsection{Materials}
We prepared the soft substrates by mixing polydimethylsiloxane (PDMS) polymers (Gelest, DMS-V31) with dimethylsiloxane copolymer crosslinkers (Gelest, HMS-301). The resulting mixtures were cured at 40$^\circ$C for 24 hours to yield soft silicone gels. The elasticity of these gels was quantitatively controlled by the weight ratio of the crosslinkers ($k$)~\cite{Zhao2022}. By systematically varying the crosslinking density from $k=0.67$~\% to 5.0~\%,  the Young's modulus of the cured gels was set to range between $E=0.16$~kPa and $175$~kPa.

To create a soft gradient surface, we first fabricated a stiff gel film with a surface area of 5~cm $\times$ 5~cm and a thickness of $h=500$~$\mu$m. By maintaining $k\geq2.5$~\%, the Young's modulus of the stiff film was designed to be $E_2 > 80$~kPa. Subsequently, a 2~cm $\times$1~cm rectangular piece was cut from the center of the gel film and replaced with a PDMS mixture containing a low density of crosslinkers ($k \leq 1.0~\%$). The resulting substrate comprised a soft region with a Young's modulus of $E_1< 6$~kPa, surrounded by a stiff region.  To ensure that the entire substrate was flat, we precisely controlled the amount of added PDMS mixture by monitoring the interface between the soft and stiff regions using a high-resolution digital camera. Due to swelling effects, a 5~$\mu$m deformed region inevitably formed near the boundary between soft and stiff gels. Throughout the experiments, these boundary defects remained substantially smaller than the droplet sizes.

\begin{figure}
    \centering
    \includegraphics[width = 7.5cm]{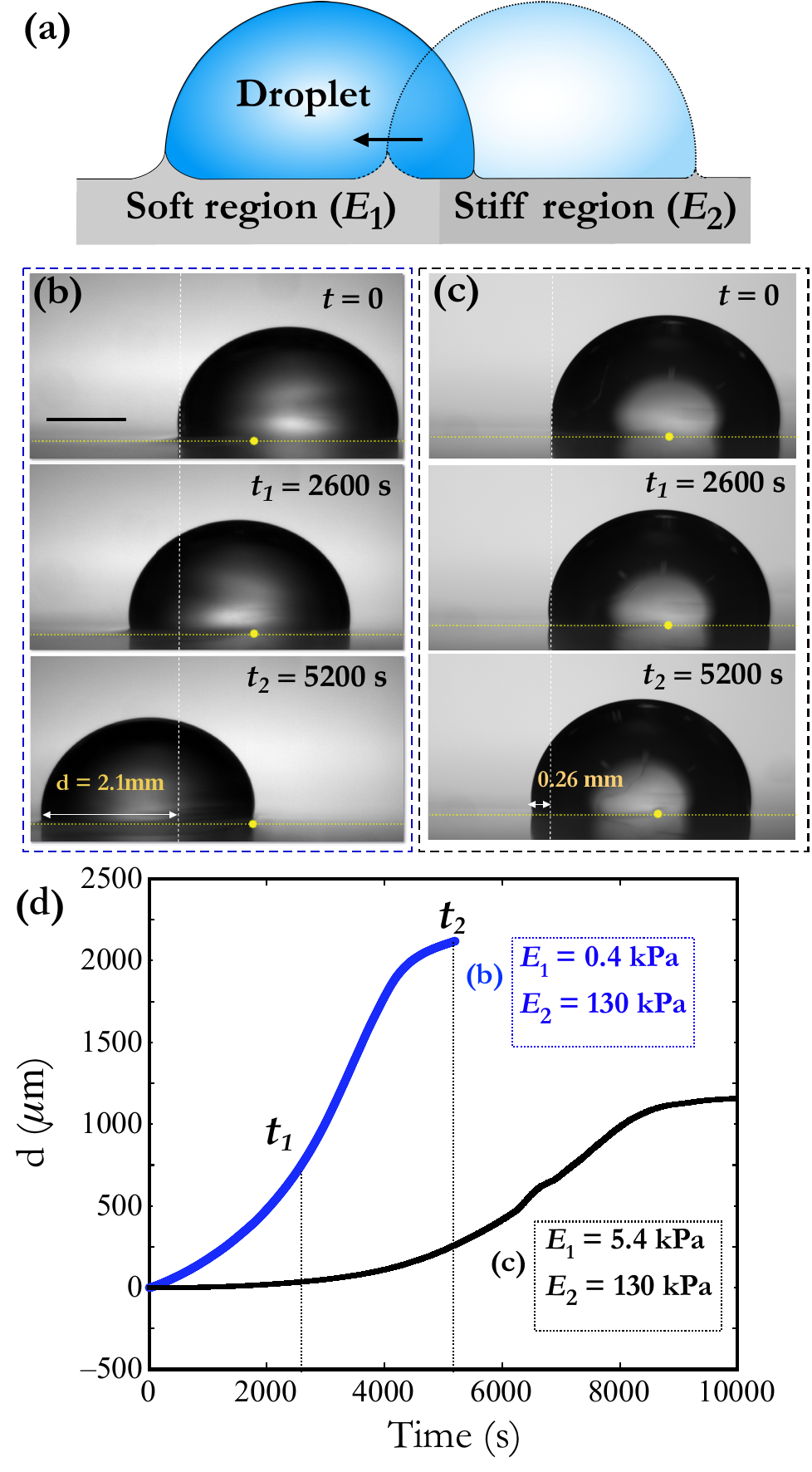}
    \caption{{\bf Droplet migration on soft gradient substrates. } {\bf (a)} Schematic illustration of a droplet moving from a stiff region ($E_2$) to a soft region ($E_1$) along a substrate. {\bf (b)} Snapshots of a moving droplet at different times: $t =0$~s, $t_1 = 2600$~s, and $t_2=5200$~s. The Young's moduli of the soft and stiff regions were $E_1 = 0.4$~kPa and $E_2 = 130$~kPa, respectively (also see Supplementary Video 1). {\bf (c)} Droplet moving on a  soft gradient substrate with $E_1 = 5.4$~kPa and $E_2=130$~kPa. {\bf (d)} Plots of droplet displacement ($d$) over time for the two droplets shown in the panels (b) and (c).}
    \label{fig:motion}
\end{figure}

\subsection{Spontaneous droplet motion}

The droplets were composed of an aqueous mixture comprising $80$~\% glycerol and $20$~\% water, with an effective viscosity of $\eta_l=48.3 \pm 1.1$~cSt. As illustrated in Fig.~\ref{fig:motion}(a), a millimeter-sized liquid droplet was initially placed on the boundary between the soft and stiff gels ($t=0$~s). Using two digital cameras, we simultaneously captured the side and bottom views of the droplet movement at $t>0$~s. The effective surface tension of these aqueous droplets on silicone gels, measured using the sessile droplet method~\cite{Zhao2022}, was $\gamma_l=41 \pm 1$~mN/m, significantly lower than the surface tension of glycerol-water mixture in air ($\sim 67$~mN/m; see Fig.~\ref{fig:mixture} in Appendix A). This difference is caused by the extraction of free chains at contact lines~\cite{Xu2020,Qian2024}. The environmental temperature and humidity were maintained at a constant $21.4 \pm 0.4$~$^\circ$C and 55.7$\pm$1.1$~\%$, respectively. Consequently, the surface tension of the water-glycerol droplet remained unchanged during the measurement period.

Figures~\ref{fig:motion}(b) and (c) show the spontaneous movements of an aqueous droplet (diameter $D = 3.5$~mm) on substrates with different combinations of $E_1$ and $E_2$. The yellow solid dots in the snapshots indicate the boundaries between gels. In both cases, the stiff regions exhibited a constant 
$E_2 =130$~kPa, while  $E_1$ of the soft region was varied from 0.4~kPa in panel (b) to $5.4$~kPa in panel (c). The droplets on both substrates spontaneously moved along the gradient of crosslinking density, with the difference between $E_1$ and $E_2$ considerably influencing the speed. Over 5200 s, the droplet in Fig.~\ref{fig:motion}(b) moved a distance of $d=2.1$~mm, while the droplet in (c)  moved only $d=0.26$~mm. Figure ~\ref{fig:motion}(d) quantitatively compares the displacements of these two droplets over time.

\section{Role of elastocapillarity}

\begin{figure}
    \centering
    \includegraphics[width = 8.5cm]{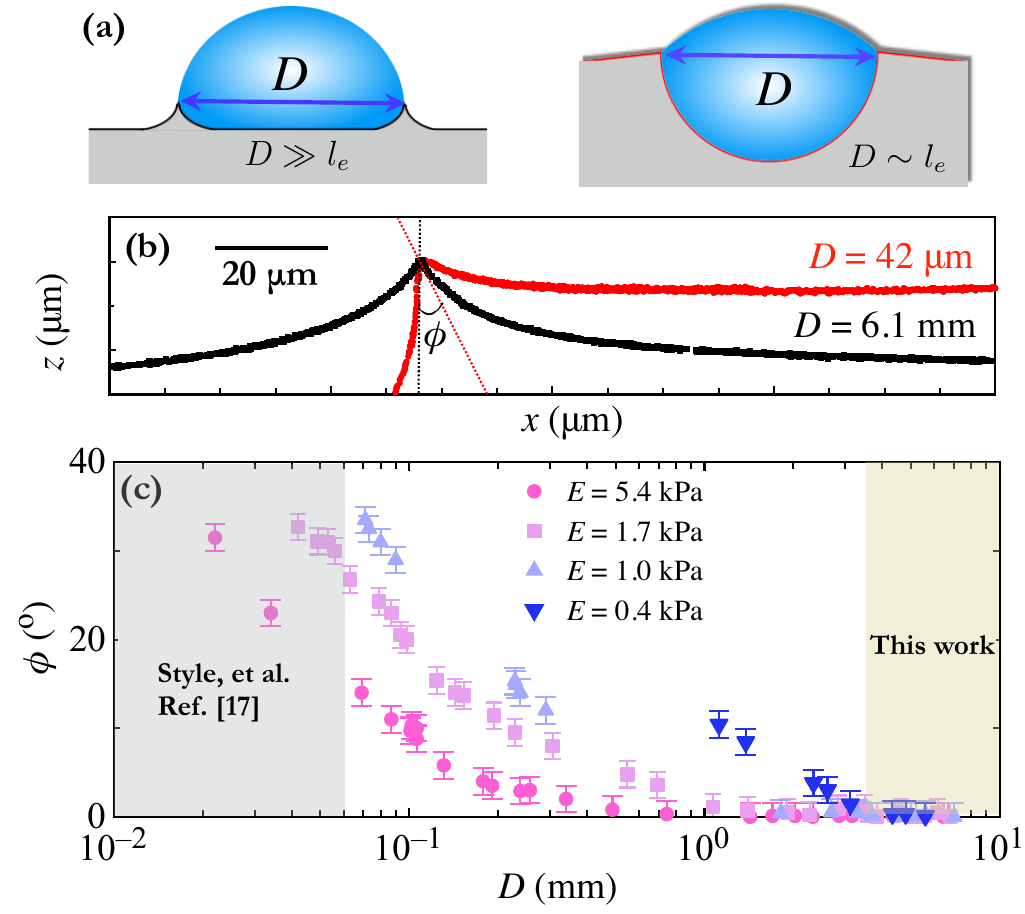}
    \caption{{\bf Role of elastocapillarity.} {\bf (a)} Schematics illustrating the rotation of the liquid--air contact line when the droplet diameter  ($D$) is comparable to the elastocapillary length ($l_e$). {\bf (b)} The red and black curves represent the wetting profiles induced by droplets with $D=42$~$\mu$m and $D=6.1$~mm, respectively, on a soft gel with $E=1.7$~kPa. The elastocapillary length ($l_e$) was approximately $10$~$\mu$m. The rotational angle ($\phi$) was determined by comparing the two profiles. {\bf (c)} Plots of $\phi$ against $D$ for $E$ = 5.4~kPa, $E$ = 1.7~kPa, $E$ = 1.0~kPa and $E$ = 0.4~kPa. The shaded gray and brown--gray areas indicate the droplet sizes investigated in Ref.~\cite{style2013patterning} and this study, respectively.}
    \label{fig:Rotation}
\end{figure}

We first examined the influence of the elastocapillary effects of soft substrates on the observed droplet movements. In general, this effect is characterized by the elastocapillary length, $l_e = \Upsilon_s/E$, where $\Upsilon_s$ is the surface stress of a soft substrate~\cite{Style2013}. As illustrated in the schematic in Fig.~\ref{fig:Rotation}(a), when the droplet diameter is comparable to the elasto-capillary length, $D\sim l_e$, the contact line rotates by an angle $\phi$ relative to the non-rotating contact line of a large droplet, $D\gg l_e$~\cite{Style2013}. Considering that $\phi$ depends on the substrate thickness, Style et al.~\cite{style2013patterning} demonstrated that micro-droplets ($D\sim l_e$) spontaneously moved along soft substrates with a thickness gradient to minimize the apparent contact angle. 

In the present study, as the droplets were several millimeters in size, the elastocapilllary effects were unlikely to play a significant role in controlling the droplet motion. To confirm this assumption, we obtained confocal images of static wetting profiles on soft gels with uniform Young's moduli ranging from $E=0.4$~kPa to 5.4~kPa. By depositing a layer of 200~nm fluorescent particles at soft interfaces, we obtained the wetting profiles using a  particle locating method~\cite{Xu2017}. For example, Fig.~\ref{fig:Rotation}(b) shows the wetting profiles induced by droplets with sizes of $D = 41$~$\mu$m and $D = 6$~mm, respectively, on the same soft gel with $E = 1.7$~kPa. As $l_e$ was approximately 15~$\mu$m, only the smaller droplet ($D = 41$~$\mu$m) exhibited a significant rotation of the wetting ridge. The rotational angle $\phi$ of the contact line was quantified by comparing the two ridge profiles. Figure~\ref{fig:Rotation}(c) shows the plots of $\phi$ against $D$ on soft gels with different Young's moduli ($E=0.45$~kPa, 0.7~kPa, 1.5~kPa, and 5.4~kPa) and a constant sample thickness of $h=500$~$\mu$m. For each $E$, $\phi$ initially decreased with $D$ until it became negligible for large droplets. The critical droplet size that marked the crossover between the two regimes was characterized by $l_e = \Upsilon_s/E$ and thus decreased with $E$. Across all the tested substrates, we consistently observed that $\phi \approx 0^\circ$ as $D>3.3$~mm.  Hence, for the millimeter-sized droplets in this study, we concluded that elastocapillarity played an insignificant role in driving the movements of the millimeter-sized droplets examined in this study.

\section{ Forces driving droplet movement}
 \subsection{Analysis of droplet dynamics}             
\begin{figure}
    \centering
    \includegraphics[width = 8.5cm]{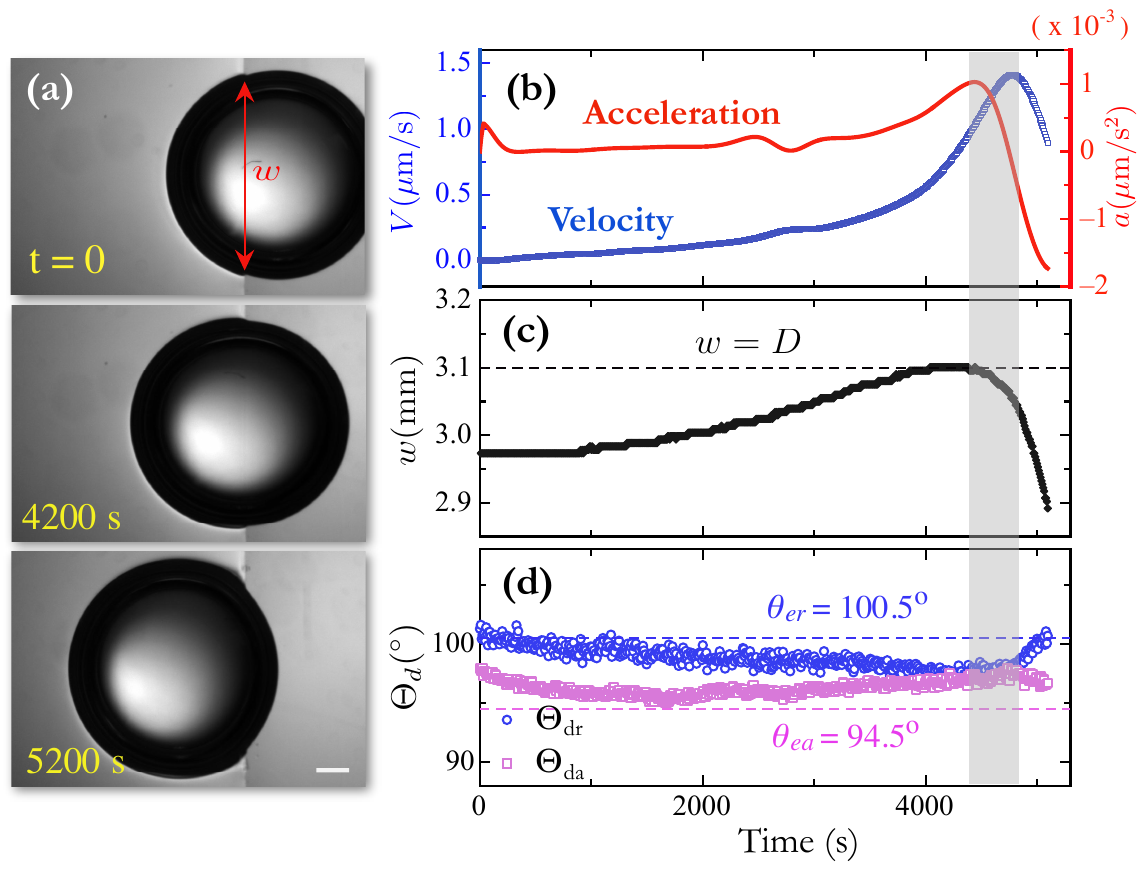}
    \caption{{\bf Dynamics of the moving droplets.} {\bf (a)} Bottom-view snapshots of a moving droplet at $t=0$~s, 4200~s, and 5200~s. The soft region (left) and the stiff region (right) had the Young's moduli of $E_1=0.4$~kPa and $E_2=130$~kPa, respectively. The red line with arrows indicates the width $w$. Scale bar: 500~$\mu$m {\bf (b)} Plots of velocity ($V$) and acceleration ($a$) against time for the droplet in (a). {\bf (c)} Plot of $w$ against time. {\bf (d)} Plots of the advancing dynamic contact angle $\Theta_{da}$ (open blue circles) and receding dynamic contact angle $\Theta_{dr}$ (pink open squares) against time. The blue and pink dashed lines indicate the equilibrium contact angles in the advancing ($\theta_{ea}$) and receding directions ($\theta_{er}$), respectively. The gray region across (b)--(d) indicates a period when the acceleration ($a$) and the velocity ($V$) of the droplet reach their peaks almost simultaneously. Within this period, $w = D$ and $\Theta_{dr}=\Theta_{da}$.
     }
    \label{fig:analysis}
\end{figure}

To explore the mechanisms underlying the spontaneous droplet motion, we assessed the macroscopic shapes of the moving droplets.  Figure~\ref{fig:analysis}(a) shows the snapshots of a liquid droplet moving from a stiff region ($E_2 = 130$~kPa) to a soft region ($E_1 = 0.4$~kPa) over a period of 5200~s. The center of this droplet was positioned in the stiff region at $t=0$~s. Subsequently, the droplet started moving toward the soft region before stopping after crossing the gel boundary. Figure~\ref{fig:analysis}(b) shows the plots of velocity $V(t)$ and acceleration $a(t)$ of the droplet center over time, where the peaks of $V$ and $a$ appear close together. 

We also characterized the width ($w$) of the droplet intersecting the interface between the two gels. Figure~\ref{fig:analysis}(c) shows the plot of $w$ against time, with the maximum width corresponding to the droplet diameter, $w = D$. A comparison of Figs.~\ref{fig:analysis}(b) and (c) shows that both $a$ and $w$ reached their maximum values simultaneously.  To validate the generality of this observation, we systematically varied both the droplet size ($D$) and the difference in the Young's moduli between the two gels ($E_1$ and $E_2$). Across all measurements, the peak values of $a$ and  $w$ were synchronized in time (Fig.~\ref{fig:add_fig} in the Appendix B). This result indicates that the maximum force propelling a droplet consistently occurs when the center of the droplet  reaches the boundary of the two gels.

We herein consider the unbalanced capillary forces acting on the dynamic contact lines. Previous studies on solid–-liquid friction have revealed the quantitative relationship between the contact line force and dynamic contact angles~\cite{Gao2018, li2023kinetic}. At the advancing contact line of a moving droplet, the net lateral capillary force can be expressed as
\begin{equation}
F_a = \kappa \gamma_{l} w (\cos\theta_{ea}-\cos\Theta_{da}),
\end{equation}
where $\theta_{ea}$ and $\Theta_{da}$ denote the static and dynamic contact angles on the soft regions, respectively; and $\kappa \approx 1.1$ is a geometrical constant for aqueous droplets wetting silicone gels~\cite{Gao2018}. Similarly, the lateral capillary force at the receding contact line can be written as
\begin{equation}
F_r = \kappa \gamma_{l} w (\cos\Theta_{dr}-\cos\theta_{er}), 
\end{equation}
where $\theta_{er}$ and $\Theta_{dr}$ denote the static and dynamic contact angles on the stiff regions, respectively. Therefore, the net contact force acting on the droplet becomes
\begin{align}
\notag F_d & = F_a + F_r \\
& = \kappa \gamma_l \omega [(\cos\Theta_{dr}-\cos\Theta_{da}) + (\cos\theta_{ea}-\cos\theta_{er})].
\label{eqn:driving}
\end{align}
In Eq.~\ref{eqn:driving}, the term in the first parenthesis represents the difference between the advancing and receding contact angles ($\Theta_{da}$ and $\Theta_{dr}$, respectively), while the term in the second parenthesis represents the difference between the equilibrium contact angles on soft and stiff gels ($\theta_{ea}$ and $\theta_{ra}$, respectively).

To estimate $F_d$, we measured the dynamic contact angles using a side camera.  Figure~\ref{fig:analysis}(d) exhibits the plots of $\Theta_{da}$ (red dots) and $\Theta_{dr} $ (blue circles) over time for the droplet shown in (a), with the solid red line and blue dashed line representing the equilibrium contact angles, $\theta_{ea} = 94.5^\circ $ and $\theta_{er} = 100.5^\circ$, respectively. Throughout the measurement period, $\theta_{ea} \leq \Theta_{da}$ and $\theta_{er} \geq \Theta_{dr}$, and thus, both contact forces $F_a$ and $F_r$ remained positive during the droplet movement. We found that $\Theta_{da} \approx \Theta_{dr}\approx 97.5^\circ$ when $w=D$, such that Eq.~\ref{eqn:driving} yields the maximum driving force  
\begin{equation}
F_{dmax} = \kappa \gamma_l D (\cos\theta_{ea}-\cos\theta_{er}).
\label{eqn:driving_max}
\end{equation} 
As $\Theta_{da} \approx \Theta_{ra}$ consistently occurred when the droplet centers reached the boundary between the two regions throughout our experiments (Fig.~\ref{fig:add_fig} in Appendix B), Eq.~\ref{eqn:driving_max} suggests that $F_{dmax}$ was determined by the difference between $\theta_{ea}$ and $\theta_{er}$. 

Additionally, researchers have revealed that the slight permeability of PDMS may drive microfludic flows on the material surfaces~\cite{Randall2005}. However, this driving mechanism was excluded in the present study due to the large scales of the moving droplets.

 \begin{figure}
    \centering
    \includegraphics[width = 8.5cm]{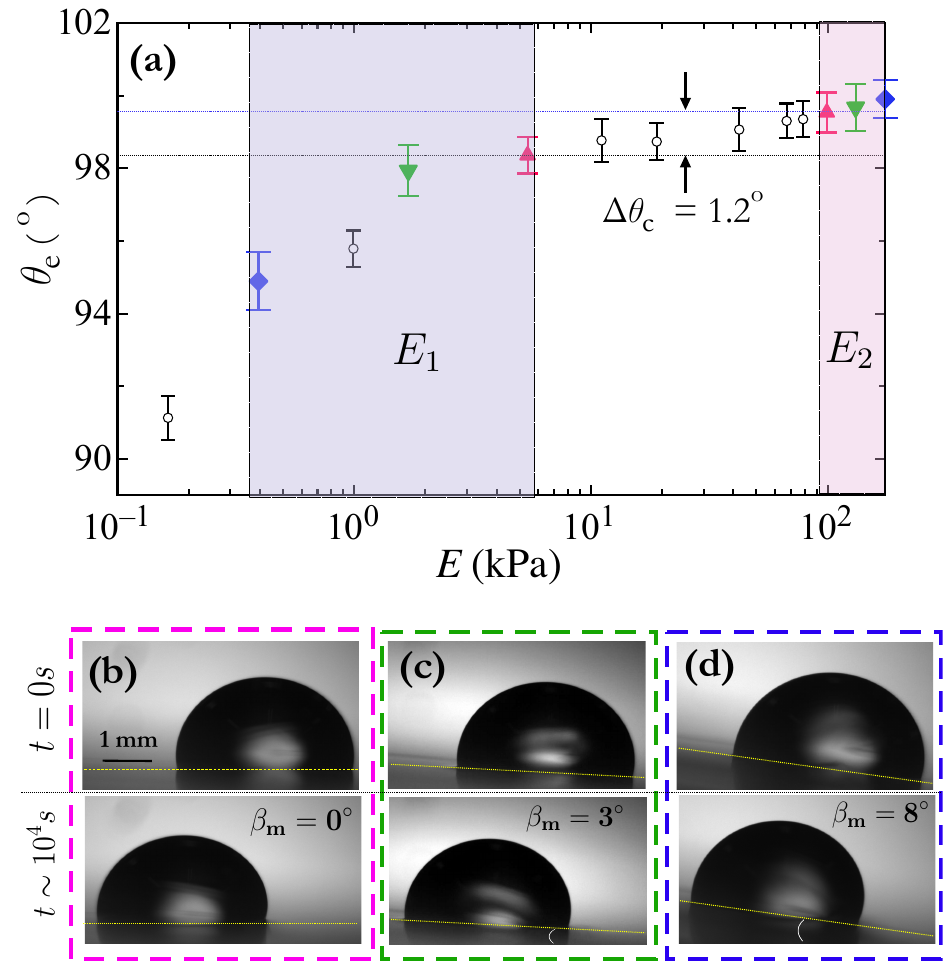}
    \caption{{\bf Spontaneous motion of droplets against gravity.} {\bf (a)} Plot of the equilibrium contact angle ($\theta_e$) of a 6-mm static droplet against the Young's modulus of the substrate ($E$). The substrate thickness was consistently $h=500~\mu$m. The two shaded areas represent the range of selected $E_1$ and $E_2$, respectively. The paired data points with different colors (red up-triangles, green down-triangles, and blue diamonds) indicate the selected $E_1$ and $E_2$ of the substrates, the droplet movements on which are illustrated in the following panels. {\bf (b)} With $E_1 = 5.4$~kPa and $E_2= 99$~kPa, a 3.5-mm droplet can only move on a level substrate ($\beta_{m}=0^\circ$). {\bf (c)} With $E_1=1.7$~kPa and $E_2=130$~kPa, a 3.5-mm droplet can climb a plane with a maximum tilted angle of $\beta_{m}=3^\circ$ (see Supplementary Video 2). {\bf (d)} With $E_1=0.4$~kPa and $E_2=175$~kPa, a 3.5-mm droplet can climb a plane with a maximum tilted angle of $\beta_{m}=8^\circ$. The yellow dotted lines represent the slope surfaces.}
    \label{fig:AnGmotion}
\end{figure}

\subsection{Criterion of spontaneous droplet motion }

To understand how the difference between $\theta_{ea}$ and $\theta_{er}$ controlled droplet motion, we varied the selections of $E_1$ and $E_2$ based on the plot in Fig.~\ref{fig:AnGmotion}(a), which shows the equilibrium contact angle ($\theta_e$)  of a millimeter-sized static droplet wetting silicone gels with different Young's moduli ($E$). Given the negligible elastocapillary effects for these large droplets, the relationship between $\theta_e$ and $E$ indicates an increase in surface energy of silicone gels with crosslinking density. The selected ranges of $E_1$ and $E_2$ are indicated by the two gray-shaded regions in panel (a). We found that a minimum difference between the equilibrium contact angles in the two gel regions, $\Delta \theta_c = \theta_{er} - \theta_{ea} = 1.2^\circ$, was required to initiate spontaneous droplet motion. 

To illustrate the tendency of these spontaneous movements, we analyzed how droplets climbed inclined surfaces against gravity.  Figures~\ref{fig:AnGmotion}(b)–(d) demonstrate the steepest slopes, characterized by a tilted angle $\beta_m$, for a droplet to climb spontaneously with different selections of $E_1$ and $E_2$. For $E_1 =5.4$~kPa and $E_2 =99$~kPa (pink up-triangles in panel (a)),  $\theta_{er} - \theta_{ea}$ is approximately $1.2^\circ $, which equals $\Delta\theta_c$. Within experimental uncertainties, a 3.5-mm droplet could move only on a level substrate, $\beta_{m} = 0^\circ$ (panel (b)). For $E_1=1.7$~kPa and $E_2 = 130$~kPa (green down-triangles in panel (a)), $\theta_{er} - \theta_{ea} \approx 1.8^\circ  >\Delta \theta_c$, allowing a droplet to spontaneously climb a tilted plane with a maximum angle of $\beta_{m} = 3.0^\circ \pm 0.5^\circ$ (panel (c)). By further increasing the stiffness gradient by using $E_1 = 0.4$~kPa and $E_2=175$~kPa (blue diamonds in panel (a)), the difference in equilibrium contact angle became as large as $\theta_{er} - \theta_{ea} \approx 5.2^\circ  >\Delta \theta_c$. Consequently, a droplet of the same size moved on a slope with $\beta_{m} = 8.0^\circ \pm 0.6 ^\circ$ (panel (d)). 

\begin{figure}
	\centering
	\includegraphics[width = 7cm]{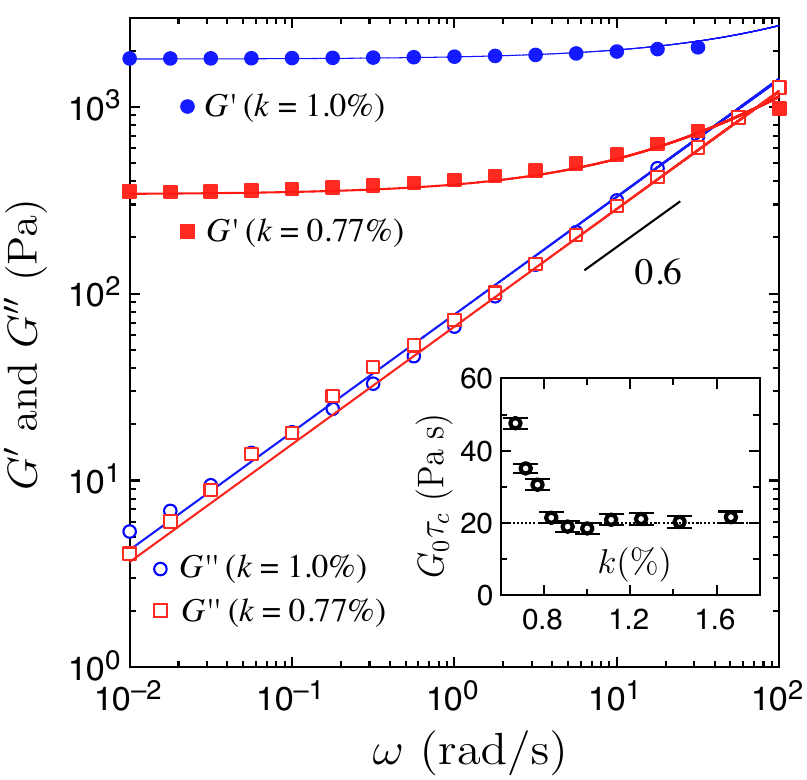}
	\caption{\textbf{Rheological properties of soft gels.} Plots of $G^\prime(\omega)$ and $G^{\prime\prime}(\omega)$ measured at a strain amplitude $\gamma_0 = 1.0~\%$ for two different crosslinking densities: $k=1.0~\%$ and $0.77~\%$. Solid lines indicate the best fits to the Chasset--Thrion model defined in Eq.~\ref{eqn:powerlaw}. Inset: plot of the fitted $G_0 \tau_c$ against $k$.}
	\label{fig:power-law}
\end{figure}

\section{Dissipation from the substrates}

\subsection{Viscoelastic rheology of soft silicone gels}
As the droplets consistently stopped shortly after crossing the boundary between the soft and stiff regions, significant dissipations from the substrates were expected.  We herein examine the viscoelastic contribution from the moving wetting ridges at contact lines~\cite{Shanahan1995, Karpitschka2015,Zhao2018}. The complex moduli of soft silicone gels are often described using the Chasset--Thirion model~\cite{Zhao2022}: 
\begin{equation}
G^*(\omega) = G^\prime (\omega) + i G^{\prime\prime} (\omega) = G_0 [\ 1 + (i \omega \tau_c)^n],
\label{eqn:powerlaw}
\end{equation}
where $G_0$ is the linear shear modulus at zero frequency, and $\tau_c$ is a characteristic relaxation timescale. Hence, the Young's modulus is given by $E=2(1+\nu)G_0$, with a Poisson ratio of $\nu \approx 0.46$ for our silicone gels~\cite{Zhao2022}. The scaling index $n$ in Eq.~\ref{eqn:powerlaw} typically ranges between $1/2$ and $2/3$ for covalently crosslinked polymer networks~\cite{Scanian1991}. 

Figure~\ref{fig:power-law} shows the plots of the storage modulus $G^\prime (\omega)$ and loss modulus $G^{\prime\prime}(\omega)$ of the prepared soft silicone gels with two different crosslinking densities, $k=0.77~\%$ and $1.0~\%$, respectively.  The solid lines represent the best fits to the Chasset--Thirion model (Eq.~\ref{eqn:powerlaw}) with $G_0 = 1.84$~kPa and $\tau_c = 11$~ms for $k=1.0~\%$, and $G_0 = 340$~Pa and $\tau_c = 91$~ms for $k=0.77~\%$, while $n = 0.61$ for both crosslinking densities. By systematically varying the crosslinking density from $k = 0.67$~\% to $k =1.67$~\%, we found that the fitted product of $G_0\tau_c$ remained constant at approximately $20$~Pa when $k>0.8\%$, as shown in the inset of Fig.~\ref{fig:power-law}. When $k > 1.7~\%$ or $E>80$~kPa,  fitting the complex moduli to the Chasset-Thirion model becomes difficult due to the short relaxation time, $\tau_c <8$~ms. As a result, 

Considering a typical dissipation frequency $\omega_0 = V/l_e = VG_0/\gamma_l$ and a characteristic deformation strain $\epsilon = \gamma \sin{\Theta}/\gamma_l$ with $\Theta$ being the dynamic contact angle, the power per unit length dissipated by a moving contact ridge can be approximated as $P_{diss} \propto G_0 (\omega_0\tau_c)^n \epsilon^2 \omega_0$ in the quasi-static limit ($V \ll \gamma_l/(G_0\tau_c)$)~\cite{Karpitschka2015, Zhao2018}. Therefore, the dissipation power explicitly becomes
\begin{align}
\notag P_{diss}  &\propto G_0 (\frac{\tau_c V}{l_e})^{n} V l_e (\frac{\gamma_l}{\Upsilon_{s}}\sin{\Theta})^{2} \\
& \approx \Upsilon_{s}V (\frac{\tau_c G_{0}V}{\Upsilon_{s}})^{n}(\frac{\gamma_l}{\Upsilon_{s}}\sin{\Theta})^{2}.
\label{eqn:dis}
\end{align} 
The resulting dissipative force per unit length is $f_{diss} = P_{diss}/V$.

\begin{figure*}
    \centering
    \includegraphics[width = 17cm]{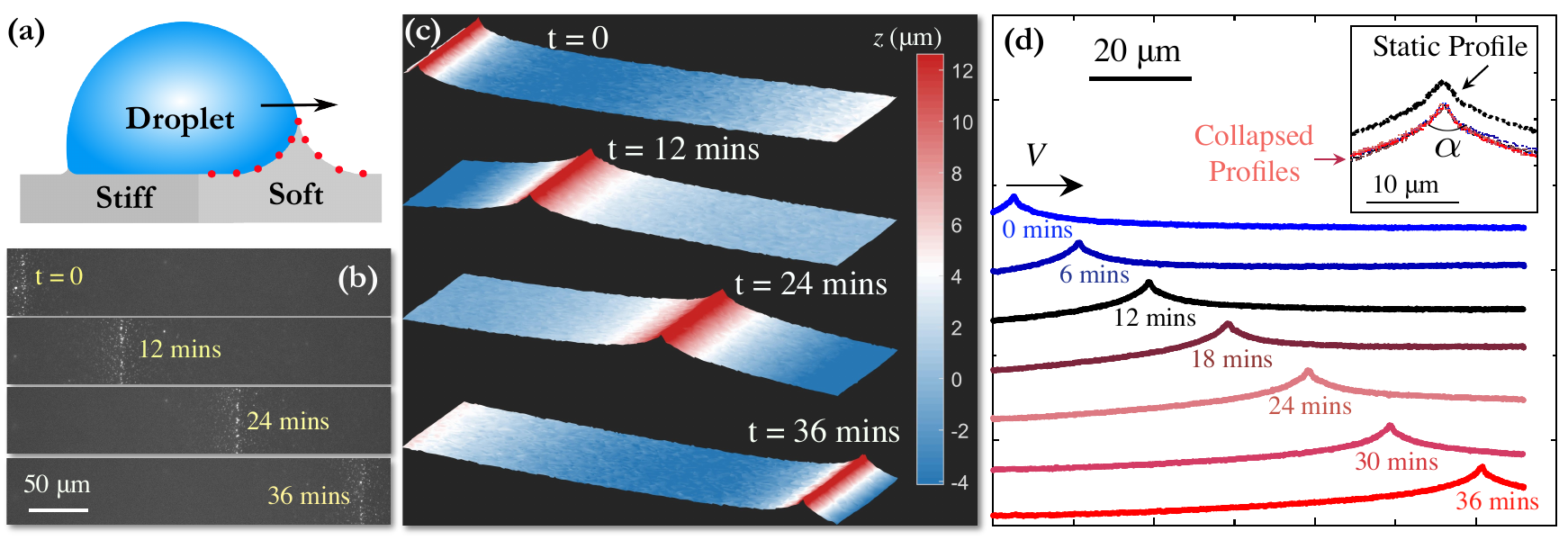}
    \caption{\textbf{Profiles of a moving contact ridge.} {\bf (a)} Schematic illustration of the confocal microscopy setup for measuring the moving contact profile in the advancing direction. The red dots represent the 200-nm fluorescent nano-beads. {\bf (b)} Fluorescence images of a contact line advancing along a substrate with $E_1=5.4$~kPa and $E_2=130$~kPa. {\bf (c)} Reconstructed three-dimensional (3D) images of the advancing contact ridge (on the soft region with $E_1=5.4$~kPa) at $t=0~$min, 12~min, 24~min, and 36~min. {\bf (d)}  Two-dimensional (2D) plots of the advancing contact ridge at different times. The plots are shifted vertically for visualization. Inset: Profiles shown in (d) collapsed near the contact point and compared with the static wetting profile (the black dots) measured over soft gels with $E=5.4$~kPa.  }
    \label{fig:DynamicMotion}
\end{figure*}

To validate the quasi-static condition assumed in Eq.~\ref{eqn:dis}, we used confocal microscopy to image the advancing contact ridges on the soft region, when the droplet centers approached near the boundary between two gels ($w \approx D$). Conversely, the receding contact ridges within the stiff region ($E_2 > 80$~kPa) were significantly smaller than 1~$\mu$m, making them difficult to be  resolved optically (Fig.~\ref{fig:DynamicMotion}(a)). The fluorescent snapshots in Fig.~\ref{fig:DynamicMotion}(b) illustrates how an advancing contact ridge moved on a substrate with $E_1 = 5.4$~kPa and $E_2 = 130$~kPa. Given the difference in equilibrium contact angle between the two regions $\theta_{ea} -\theta_{er} \approx 1.3^\circ \approx \Delta \theta_c$, the droplet moved very slowly ($\sim 0.5~\mu$m/s), such that a three-dimensional (3D) confocal scanning was properly implemented in-situ. As shown in Fig.~\ref{fig:DynamicMotion}(c), a 3D particle locating method was employed to reconstruct the moving ridge profiles over a period of 36~mins, which were further collapsed to two-dimensional (2D) plots in Fig.~\ref{fig:DynamicMotion}(d) by considering the azimuthal symmetry. Since the advancing speed was significantly lower than $\gamma_l/(G_0\tau_c) \sim 10$~$\mu$m/s, the quasi-static assumption of Eq.~\ref{eqn:dis} remained valid. The inset in panel (d) further verifies that the local geometries of the moving ridges were consistent with the static profile, featuring a constant opening angle of $\alpha = 68^\circ \pm 1.3^\circ$. Using the Neumann triangle for local stress balance~\cite{Style2013}, we obtained the surface stress of the soft region as $\Upsilon_s = 2\gamma_l/\cos{(\alpha/2)} = 25.3 \pm 0.6$~mN/m. 

\subsection{Estimation of the dissipative forces}
When the droplet centers arrived at the boundary between two gels ($w = D$), the driving force $F_d$ reached its maximum value $F_{dmax}$ given by Eq.~\ref{eqn:driving_max}. Meanwhile, the maximum dissipative force $F_{dmax}$ due to viscoelasticity of soft gels can be obtained by integrating $f_{diss}$ along the contact perimeter:
\begin{equation}
F_{diss} = f_{diss}\cdot D \approx \Upsilon_{s}D (\frac{\tau_c G_{0}V}{\Upsilon_{s}})^{n}(\frac{\gamma_l}{\Upsilon_{s}}\sin{\Theta})^{2}.
\label{eqn:dissipation}
\end{equation}

We next estimated the magnitudes of $F_{dmax}$ and $F_{diss}$ for the moving droplet shown in Fig.~\ref{fig:DynamicMotion}. First, using $D=3.2$~mm, $\theta_{ea} = 99.7^\circ$ and $\theta_{er} = 98.4^\circ$, we obtained that $F_{dmax} \approx 3.2$~$\mu$N according to Eq.~\ref{eqn:driving_max}. Second, to estimate of $F_{diss}$, the dissipations of moving ridges on both soft and stiff regions were considered. When $w=D$, we found that $\Theta_{dr}\approx\Theta_{da} \approx 99^\circ$ and $V\approx 0.5$~$\mu$m/s. Together with the viscoelastic parameters fitted for $k=0.77~\%$  in Figs.~\ref{fig:power-law}, the dissipative force at the advancing contact line (the soft region) was estimated as $F_{diss1} \approx 1.5$~$\mu$N using Eq.~\ref{eqn:dissipation}. As the complex moduli and surface stresses of stiff gels ($E_2 >80$~kPa) were challenging to determine experimentally, we  estimated only the upper bound of the dissipative force in the receding direction. Given the increase in surface stress with crosslinking density~\cite{Zhao2022}, we anticipated that $\Upsilon_s > 35$~mN/m for $E_2 = 130$~kPa. Using $G_0 \tau_c \sim 20$~Pa from Fig.~\ref{fig:power-law},  the dissipative force at the receding contact line was estimated to be $F_{diss2} < 1.0$~$\mu$N by Eq.~\ref{eqn:dissipation}. Consequently, the total dissipative force induced by the moving wetting ridges became $F_{diss} = F_{diss1} + F_{diss2} <  2.5~\mu$N. Considering that both Eqs.~\ref{eqn:driving_max} and \ref{eqn:dissipation}  serve as only the estimations of relevant force scales, we conclude that $F_{diss}$ remains approximately close to or slightly lower than $F_{dmax}$. By varying the combinations of $E_1$ and $E_2$, we found that the relationship $F_{diss} \lesssim F_{dmax}$ remained valid throughout our experiments (Appendix B).

Further, other dissipative factors may have contributed to decelerating the droplets. Given the low capillary number of the moving droplets, $Ca = \eta_l V/\gamma_l \sim 10^{-7}$, the hydrodynamic dissipation within the moving droplets is negligible~\cite{li2023kinetic,li2022spontaneous}. However, since it is challenging to prevent surface deformations at the interface between two gels, we anticipate that the boundary could act as a pinning defect, dampening the droplet motion~\cite{Guan2016}. 

\section{Conclusions}

{\color{red}}

We experimentally observed spontaneous droplet motion on the surfaces of soft gels with gradients in crosslinking density (Fig.~\ref{fig:motion}). These movements were driven by the difference in equilibrium contact angles between the advancing and receding directions (Fig.~\ref{fig:analysis}). As the droplets traversed the boundary between gels, the unbalanced capillary force overcame the dissipations from the substrates (Fig.~\ref{fig:DynamicMotion}). We further demonstrated that increasing the difference of crosslinking density allowed droplets to climb a tilted plane with a steep slope (Fig.~\ref{fig:AnGmotion}).

As the moving droplets were at least a few millimeters in size, we concluded that the elastocapillary effects of the soft substrates, whose influences on the motion of micron-sized droplets were previously investigated ~\cite{style2013patterning}, played an insignificant role in this study (Fig.~\ref{fig:Rotation}). By measuring the droplet dynamics under the  condition $\theta_{er} - \theta_{ea} > \theta_{c} \approx 1.2^\circ$, we attributed the spontaneous droplet migration to the surface energy gradient resulting from the spatially varying crosslinking density~\cite{Zhao2022}. Future works could focus on exploring systems with a continuous gradient of crosslinking density, which may help prevent dissipations from boundaries between sharply changing gel matrices. Additionally,  experimental studies of such systems would facilitate comparative analyses with existing theoretical predictions~\cite{Bardall2020, bueno2018wettability,theodorakis2017stiffness}.

\section*{Supplemental Materials}
See the supplementary videos demonstrating the spontaneous movements of liquid droplets along soft gradient surfaces

\section*{Acknowledgments}
We appreciate Caishan Yan for engaging in insightful dicussions. This study was financially supported by the General Research Funds (16305821 and 16306723), the Early Career Scheme (26309620), and the Collaborative Research Fund (C6004-22Y) from the Hong Kong Research Grants Council.

\section*{Author Declarations}
\subsection*{Conflict of Interest}
The authors have no conflicts to disclose.

\subsection*{Author Contributions}
\textbf{Weiwei Zhao:} Conceptualization (equal); Investigation (lead); Methodology (equal); Writing – original draft (equal); \textbf{Wenjie Qian:} Investigation (equal); Methodology (equal); Writing – review \& editing (equal); \textbf{Chang Xu:} Investigation (equal);  Methodology (equal); \textbf{Qin Xu: }Conceptualization (lead); Investigation (equal); Writing – review \& editing (equal).

\section*{Data Availability}
The data that support the findings of this study are available from the corresponding author upon reasonable request.

\section*{Appendix A: glycerol--water mixture}
\begin{figure}
    \centering
    \includegraphics[width = 8cm]{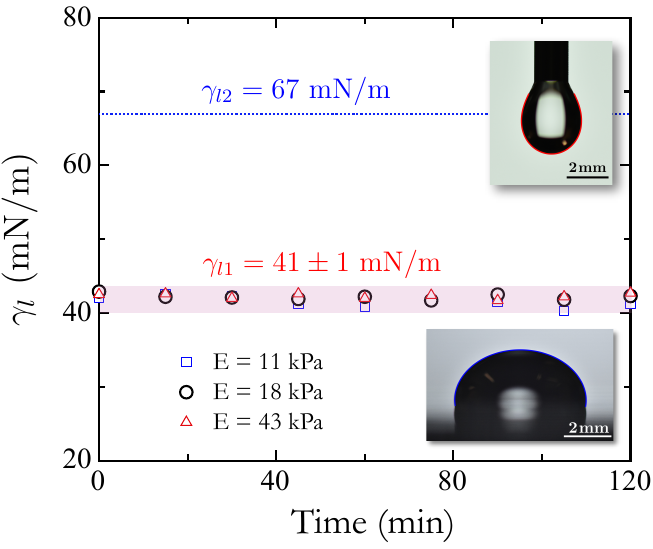}
    \caption{\textbf{Surface tension of a glycerol--water mixture.} The surface tension of the glycerol (80\%)--water (20\%) mixture in air was measured to be $\gamma_{l2}  = 67$~mN/m (blue dotted line) by using the pendant droplet method. In addition, the surface tension of the sessile droplet on silicone gels was measured to be $\gamma_{l1} = 41$~mN/m on soft gels with different Young's moduli, and this value remained unchanged over two hours.}
    \label{fig:mixture}
\end{figure}
Liquid droplets were prepared by mixing 80~\% w/w glycerol with 20~\% w/w water. To ensure the physical and chemical stabilities of the droplets, we consistently maintained the environmental temperature at 21.4 $\pm$ 0.4$^\circ$C, and the humidity at 55.7$\pm$ 1.1$~\%$. Under these measurement conditions, the surface tension of the glycerol--water mixture was characterized using both sessile and pendant droplet methods. 

The open points in Fig.~\ref{fig:mixture} show the surface tension of sessile droplets ($\gamma_{l1}$) measured at different times on soft gels with different moduli ($E$). By considering the balance between capillary pressure and gravity, we found that $\gamma_{l1} = 41 \pm 1$~mN/m regardless of $E$, and this value remained constant over a two-hour period. The finding indicates that the physical properties of the sessile droplets on soft gels remained unchanged throughout our experiments. In contrast, the surface tension of a pendent mixture droplet was measured as $\gamma_{l2} = 67$~mN/m (blue dotted line in Fig.~\ref{fig:mixture}). The difference between $\gamma_{l1}$ and $\gamma_{l2}$ was due to the migration of free chains from soft gels that covered the sessile droplets. As we focused on droplets moving on soft gel surfaces in this study, the value of $\gamma_{l1}$ was used as the effective liquid surface tension $\gamma_l$. Furthermore, the viscosity ($\eta_l$) of the glycerol--water mixture was characterized by an Anton Paar MCR 302 rheometer equipped with a cone--plate shear cell.

\section*{Appendix B: additional results of the moving droplets}

\begin{figure}
    \centering
    \includegraphics[width = 8.8cm]{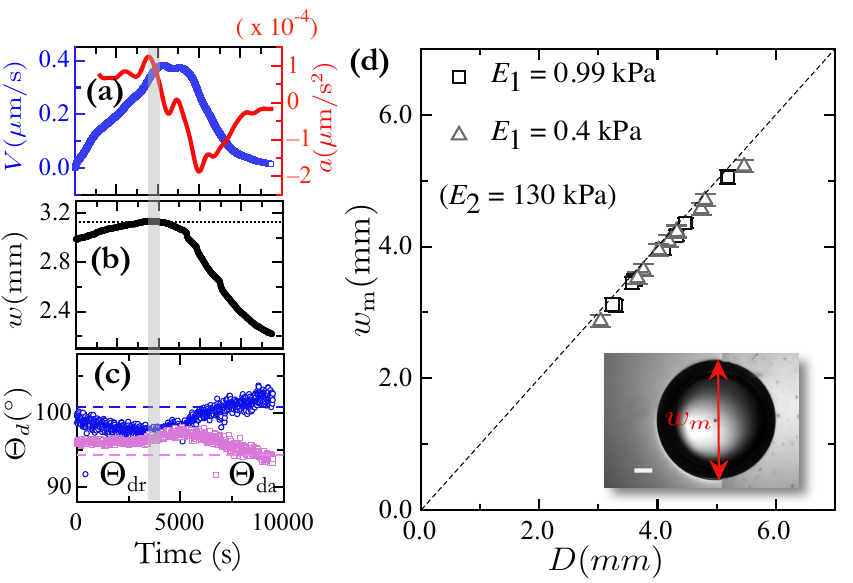}
    \caption{\textbf{Additional data of droplet dynamics.} (a) -- (d). Droplet dynamics on the soft substrates with $E_1 = 0.4$~kPa and $E_2 = 130$~kPa,  supplementing the results show in Fig.~\ref{fig:analysis}.  The velocity ($V$), the acceleration ($a$), the width intersecting the gel boundary ($w$), and the dynamic contact angle ($\Theta_d$) of the droplet are plotted against time. The dotted line in (b) indicates $\omega = D$. The dashed lines in (c) represents $\theta_{er} = 100.5^\circ$ (blue) and $\theta_{ea} = 94.5^\circ$ (pink). (d) Plot of $w_m$ against $D$ for $E_1 =0.4$~kPa and 0.99~kPa, with $E_2$ maintained constant at 130~kPa. The dotted line represents the scaling of $w_m = D$. Inset: $w_m$ is defined as the droplet width intersecting with the boundary between gels when the droplet acceleration is maximized.}
    \label{fig:add_fig}
\end{figure}

In addition to the representative results of droplet dynamics shown in Fig.~\ref{fig:analysis}, we repeated our measurements on each gel substrate for at least three different droplets. For instance, Figs.~\ref{fig:add_fig}(a)--(c) shows the plots of the velocity ($V(t)$),  acceleration ($a(t)$), the intersecting width ($w(t)$), and the dynamic contact angle ($\Theta_d(t)$) of a different droplets moving on the same soft substrate with $E_1 = 0.4$~kPa and $E_2 = 130$~kPa (see Supplementary Video 3). Similar to the findings in Fig.~\ref{fig:analysis}, we found that the acceleration peaked when the droplet center arrived at the boundary between gels, $w=D$. At this time point, the advancing and receding contact angels are approximately equal, $\Theta_{da} \approx \Theta_{dr}$. 

Moreoever, we systematically varied the droplet diameter from $D=2$~mm to 4.5~mm on the substrates with $E_1 = 0.4$~kPa and $0.99$~kPa, respectively, while maintaining $E_2 = 130$~kPa. At the moment of the maximum acceleration, the intersecting width of the droplet with the boundary between gels ($w_m$) was determined. Figure~\ref{fig:add_fig}(b) shows the plot of $w_m$ against $D$ for all the droplets, which aligns well with the scaling of $w_m = D$. This result confirms that the droplets experienced maximum acceleration when their centers reached the boundary between gels.

We further evaluate the maximum driving force $F_{dma x}$ and the dissipative force $F_{diss}$ acting on the moving droplets illustrated in Fig.~\ref{fig:analysis} and Fig.~\ref{fig:add_fig}, respectively. As $E_1 = 0.4$~kPa and $E_2 = 130$~kPa, the maximum driving force is approximated as $F_{dmax} = \kappa \gamma_l (\cos{\theta_{ea}}-\cos{\theta_{er}}) \approx 15$~$\mu$N, where $\theta_{ea} =94.5^\circ$ and $\theta_{er} = 100.5^\circ$ for both droplets. The maximum velocity reached is $V \approx 1.2$~$\mu$m/s for the droplet in Fig.~\ref{fig:analysis}, while $V \approx 0.4$~$\mu$m for the droplet in Fig.~\ref{fig:add_fig}. Using Eq.~\ref{eqn:dissipation}, we find that $F_{diss} < 10~\mu$N $<F_{dmax}$  for both moving droplets, which is consistent with the estimations presented in Sec.~IV.~B.

\bibliography{reference}

\end{document}